\documentclass[apj]{emulateapj}

\newcommand{\Lsun}{\mbox{\,$L_{\odot}$}}
\def\spose#1{\hbox to 0pt{#1\hss}}
\def\simlt{\mathrel{\spose{\lower 3pt\hbox{$\mathchar"218$}}
     \raise 2.0pt\hbox{$\mathchar"13C$}}}
\def\simgt{\mathrel{\spose{\lower 3pt\hbox{$\mathchar"218$}}
     \raise 2.0pt\hbox{$\mathchar"13E$}}}
\slugcomment{Accepted for publication in ApJ}

\usepackage{lscape}
\usepackage{natbib}
\usepackage{url}
\font\smcap=cmcsc10

\newcommand{\kms}{\,km~s$^{-1}$}

\newcommand{\nai}{Na\,{\smcap i}}
\newcommand{\caii}{Ca\,{\smcap ii}}

\newcommand{\vio}{$(V-I)_0$}
\newcommand{\ivi}{[$I_0$,\,($V-I$)$_0$]}

\newcommand{\feh}{$\rm[Fe/H]$}

\newcommand{\mv}{$\langle v\rangle$}
\newcommand{\mvsph}{$\langle v\rangle^{\rm sph}$}
\newcommand{\sigvsph}{$\sigma^{\rm sph}_v$}
\newcommand{\mvsb}{$\langle v\rangle^{\rm KCC}$}
\newcommand{\sigvsb}{$\sigma^{\rm KCC}_v$}

\shorttitle{The Dominance of Metal-Rich Stellar Streams}
\shortauthors{Gilbert et~al.}

\begin{document}
\bibliographystyle{apj}

\title{The Dominance of Metal-Rich Streams in Stellar Halos: A Comparison Between Substructure in M31 and $\Lambda$CDM Models}

\author{
Karoline~M.~Gilbert\altaffilmark{1},
Andreea~S.~Font\altaffilmark{2},
Kathryn~V.~Johnston\altaffilmark{3},
Puragra~Guhathakurta\altaffilmark{4} 
}

\email{
kgilbert@astro.washington.edu
} 

\altaffiltext{1}{Department of Astronomy, University of Washington, Box 351580, Seattle, WA, 98195-1580.}
\altaffiltext{2}{Institute for Computational Cosmology, University of Durham, Science Laboratories, South Road, Durham DH1 3LE, UK}
\altaffiltext{3}{Department of Astronomy, Columbia University, Pupin Physics Laboratory, 550 West 120th Street, New York, New York 10027}
\altaffiltext{4}{UCO/Lick Observatory, Department of Astronomy \&
Astrophysics, University of California Santa Cruz, 1156 High Street, Santa
Cruz, California 95064.}
\setcounter{footnote}{5}

%

\begin{abstract}
{

Extensive photometric and spectroscopic surveys of the Andromeda galaxy 
(M31) have discovered tidal debris features throughout M31's stellar halo.  
We present stellar kinematics and metallicities in fields with identified 
substructure from our on-going SPLASH survey of M31 red giant branch stars
with the DEIMOS spectrograph on the Keck~II 10-m telescope. Radial 
velocity criteria are used to isolate members of the 
kinematically-cold substructures.  The substructures are shown to be 
metal-rich relative to the rest of the dynamically hot stellar population in 
the fields in which they are found.  We calculate the mean metallicity and 
average surface brightness of the various kinematical components in each field, 
and show that, on average, higher surface brightness features tend to be more 
metal-rich than lower surface brightness features.  Simulations of stellar 
halo formation via accretion in a cosmological context are used to illustrate 
that the observed trend can be explained as a natural consequence of the observed 
dwarf galaxy mass-metallicity relation.  A significant spread in metallicity 
at a given surface brightness is seen in the data; we show that this is due 
to time effects, namely the variation in the time since accretion of the 
tidal streams' progenitor onto the host halo.  We show that in this theoretical 
framework a relationship between the alpha-enhancement and surface brightness of 
tidal streams is expected, which arises from the varying times of accretion 
of the progenitor satellites onto the host halo.  Thus, measurements of 
the alpha-enrichment, metallicity, and surface brightness of tidal debris 
can be used to reconstruct the luminosity and time of accretion onto the 
host halo of the progenitors of tidal streams.  
}
\end{abstract}


\setcounter{footnote}{0}

\section{Introduction}\label{sec:intro}
  
In a hierarchical universe, a large galaxy will undergo many mergers during its lifetime.  The amount and properties of accreted material observed in present-day galaxies can be used to test hierarchical formation scenarios.  Stellar halos provide an ideal environment for investigating the detailed merger history of an individual galaxy.  The sparse stellar populations of a galaxy's halo combined with long dynamical times make it possible for tidal debris features to remain identifiable in phase-space for billions of years.  In contrast, it is difficult to separate material formed in situ from accreted material in a galaxy's relatively dense disk and bulge.  

Recent sophisticated numerical and
semianalytic simulations of stellar halo formation have made great strides in 
characterizing the properties of stellar halos that are built up through the tidal 
stripping of accreted satellites \citep[e.g.,][]{johnston1996,johnston1998,helmi1999,helmi2000,bullock2001}.
\citet{bullock2005} (hereafter BJ05) used a combination of $N$-body and 
semi-analytic techniques to simulate the formation via accretion of 11 
MW-analog stellar halos in a cosmological context. These high resolution
simulations allow detailed analysis of the expected physical properties of stellar
halos composed primarily of tidal debris and their associated satellite systems 
\citep{font2006apjsats,font2006apjhalos,font2008,johnston2008}. 

The stellar halos of the Milky Way (MW) and Andromeda 
(M31) galaxies both show ample evidence of recent and on-going accretion events.  
Among the most prominent of these substructures are the Sagittarius 
stream \citep{ibata1994,majewski2003,newberg2003} and the 
Monoceros stream \citep{yanny2003,rocha-pinto2003}  
in the MW, and the giant southern stream 
\citep[GSS;][]{ibata2001nature} in M31, which likely pollutes much of M31's
spheroid interior to $\sim 20$\,--\,30~kpc \citep{fardal2007,gilbert2007,richardson2008}.

In addition to these prominent
stellar streams, recent large observational surveys are increasing
the number of known tidal debris features by pushing to lower surface brightness 
limits.  Analysis of the Sloan
Digital Sky Survey has revealed many faint tidal debris features in the MW's 
stellar halo \citep[e.g.,][]{grillmair2006,grillmair2006b,belokurov2006,belokurov2007b},
while large-scale photometric studies of M31 are revealing similarly large numbers
of tidal debris features in M31's stellar halo \citep{ibata2001nature,ibata2007}.
Spectroscopic surveys of red giant branch (RGB) stars in M31 
\citep{irwin2005,kalirai2006halo,gilbert2006} are enabling the 
discovery of tidal debris features through the kinematics of M31's stellar 
populations \citep[e.g.,][]{ibata2005,kalirai2006gss,gilbert2007,chapman2008,gilbert2009gss}. 

A classic signature
of substructure is an abnormally high
surface brightness relative to the value expected from the
observed surface brightness profile of the stellar halo.  Due to the long 
dynamical times in the sparse outer regions of galaxy halos,
tidal debris from relatively recent mergers
(within the last several Gyr) will not have had time to become
spatially well-mixed.  
The kinematical distribution of a well-mixed stellar halo population is
expected to be broad (dynamically hot) out to large distances from 
the center of the galaxy. 
In the MW and M31, the measured line-of-sight velocity dispersion is 
$\sigma_v\sim 120$ to 150~\kms\ near the galaxy's center 
\citep{battaglia2005,chapman2006,gilbert2007} and decreases to 
$\sigma_v\sim $~100~\kms\ at $R\sim 60$~kpc \citep{battaglia2005,chapman2006}. 
In contrast, tidal debris from a single accretion event should be 
dynamically cold, since the local velocity dispersion of the debris should 
decrease as it mixes along the progenitor satellite's orbit: as the stars 
spread apart, decreasing their spatial density, their density in velocity-space 
must increase in order to conserve their phase-space density and satisfy 
Liouville's theorem \citep{helmi1999}.
Hence, kinematical substructure in a survey is another signature of 
hierarchical structure formation.   A
difference in chemical abundance between stars in the kinematically hot
and cold populations or between two closely spaced fields is also evidence
of substructure, as the progenitors of recent accretion events are likely to be chemically distinct from the ensemble of satellites that were accreted at early times.  
 
We have utilized the kinematical and chemical signatures of substructure to identify multiple, 
kinematically cold tidal debris features \citep{guhathakurta2006,kalirai2006gss,gilbert2007,gilbert2009gss} 
during the course of the Spectroscopic and Photometric Landscape of Andromeda's Stellar Halo (SPLASH) Survey which includes extensive Keck/DEIMOS\footnote{Data presented herein were obtained 
at the W.\ M.\ Keck
Observatory, which is operated as a scientific partnership among the
California Institute of Technology, the University of California and the
National Aeronautics and Space Administration.  The Observatory was made
possible by the generous financial support of the W.\ M.\ Keck Foundation.
} spectroscopic survey of M31's stellar halo
\citep[e.g.,][]{guhathakurta2005,kalirai2006halo,gilbert2006,gilbert2007}.  
Kinematical signatures of substructure enable the detection of faint and diffuse
features which are not easily identified in starcount maps.  For example, 
a shelf feature \citep{gilbert2007} along M31's southeast minor axis 
($R_{\rm proj}\sim 12$\,--\,18~kpc) went unnoticed on star count maps 
\citep[in fact, the region was specifically targeted
for deep HST photometry based on its apparent smoothness;][]{brown2003}. 
The observed distribution of stars in line-of-sight velocity vs.\ position
space and its striking similarity to model predictions led to the
discovery of the SE shelf.

We can use the kinematically cold tidal debris features found in our survey 
to study general trends of tidal debris in stellar halos.
The recent advances in  
simulations of stellar halo formation have made it possible to undertake detailed comparisons of the properties of observed and simulated substructure \citep{bell2008,font2008}. 
In this contribution, we present a relationship between the surface brightness and metallicity 
of tidal debris features observed in our M31 data set, and provide a physical interpretation of the observed trend through comparisons of the observations with simulated stellar streams in the BJ05 simulations. 

The paper is organized as follows.  In \S\,\ref{sec:data} we briefly discuss the 
observations and reduction techniques and present the fields in our M31 survey in 
which substructure has been identified.  We present the kinematical and chemical abundance 
distributions of the stars in these fields as well as the surface brightness and average metallicity of the observed kinematical substructures.  The methods used in the BJ05 simulations are outlined
in \S\,\ref{sec:sims}, and the
trends of surface brightness with chemical abundance of stellar streams 
in the simulated BJ05 stellar halos are presented.  We compare the observations with the simulations and discuss the resulting physical implications in \S\,\ref{sec:comparison}.  Finally, we summarize our results in \S\,\ref{sec:sum}.

\section{Data}\label{sec:data}
\subsection{Photometric and Spectroscopic Observations}
The photometric and spectroscopic observations and data reductions employed
in our SPLASH M31 survey have been discussed in detail in previous papers 
\citep{guhathakurta2006,kalirai2006gss,gilbert2006,gilbert2007} 
and will be only briefly outlined here.  The 
observations are organized into ``fields'' in M31's stellar halo based
on the spatial proximity of the spectroscopic slitmasks (Table~\ref{table1}).  
A single photometric
pointing is large enough to support multiple, non-overlapping DEIMOS slitmask 
observations; the spectra from slitmasks based on a single photometric pointing 
are in general analyzed together and considered a field.  The 
exception to this are fields f207 and H13s, which are from a single 
photometric pointing but are treated as separate fields due to the 
fact that they are located at 
different positions along M31's GSS; the debris from
the GSS is centered at a different mean velocity in 
each of these fields.

\begin{deluxetable*}{lccrrl}
\tabletypesize{\scriptsize}
\tablecolumns{6}
\tablewidth{0pc}
\tablecaption{M31 Spectroscopic Fields with Identified Substructure.}
\tablehead{\multicolumn{1}{c}{Field} & \multicolumn{1}{c}{\#\ Spectroscopic} & \multicolumn{1}{c}{Projected}          & 
  \multicolumn{1}{c}{\#\ Stars\tablenotemark{a}} & \multicolumn{1}{c}{\#\ of M31} & Reference \\
&   \multicolumn{1}{c}{Masks}      & \multicolumn{1}{c}{Radius}     &  & 
\multicolumn{1}{c}{Stars\tablenotemark{b}} & \\ 
&     & \multicolumn{1}{c}{(kpc)} &  &  &}
\startdata
SE shelf & 4 & 12\,--\,18 & 510 & 428 & \citet{gilbert2007} \\ 
f207 & 1 & 16\,--\,19 & 57 & 49 & \citet{gilbert2009gss} \\ 
H13s & 2 & 20\,--\,23 & 212 & 178 & \citet{kalirai2006gss}, \\ 
     &   &            &     &     & \citet{gilbert2009gss}\\ 
a3 & 3 & 31\,--\,34 & 88 & 69 & \citet{guhathakurta2006}, \\
     &   &            &     &     & \citet{gilbert2009gss}\\ 
a13 & 4 & 55\,--\,63 & 109 & 44 & \citet{gilbert2009gss} \\ 
m4 & 5 & 53\,--\,60 & 154 & 53 & \citet{gilbert2009gss} \\ 
\enddata
\tablenotetext{a}{The total number of unique stellar spectra for which we recovered velocities.}
\tablenotetext{b}{The number of M31 RGB stars is defined as the number of stars
that are identified as secure and marginal M31 RGB stars by the \citet{gilbert2006}
diagnostic method.}
\label{table1}
\end{deluxetable*}


Photometry and astrometry for the SE shelf, f207, and H13s fields 
were derived from images in the $g'$ and $i'$ bands taken with the MegaCam instrument
on the 3.6-m Canada-France-Hawaii Telescope (CFHT).\footnote{MegaPrime/MegaCam is
a joint project of CFHT and CEA/DAPNIA, at the Canada-France-Hawaii Telescope
which is operated by the National Research Council of Canada, the Institut
National des Science de l'Univers of the Centre National de la Recherche
Scientifique of France, and the University of Hawaii.}  The SExtractor program
was used for object detection, photometry, and morphological classification \citep{bertin1996} and observations of Landolt photometric standard stars were used to transform
from instrumental $g'$ and $i'$ magnitudes to Johnson-Cousins $V$ and $I$ magnitudes \citep{kalirai2006gss}.  

Photometry and astrometry for the a3, a13, and m4 fields 
were derived by \citet{ostheimer2003} from images in the Washington $M$ and 
$T_2$ bands and the intermediate-width
DDO51 band taken with the Mosaic camera on the 4-m Kitt Peak National Observatory (KPNO) telescope.\footnote{Kitt Peak National 
Observatory of the National Optical Astronomy Observatory is operated by the 
Association of Universities for Research in Astronomy, Inc., under cooperative 
agreement with the National Science Foundation}  The DDO51 filter includes the
surface-gravity sensitive Mg\,$b$ and MgH stellar absorption features. 
The combination of these three filters allows photometric selection 
of stars that are likely to be M31 red giant branch (RGB) stars rather 
than MW dwarf stars along the line-of-sight to M31 \citep{majewski2005}.  The 
photometric transformation relations in \citet{majewski2000} were used to 
convert the $M$ and $T_2$ magnitudes to Johnson-Cousins $V$ and $I$ magnitudes.      

Objects were selected for spectroscopy based on their derived $I$ magnitudes and 
morphological properties.  In the outer fields (a3, a13, and m4), objects were further prioritized by
their probability of being an M31 RGB star based on the $M$, $T_2$, and DDO51 
photometry; this is vital for increasing the spectroscopic efficiency of our survey
in the sparse outer regions of M31's stellar halo.  Spectroscopic slitmasks were
observed with the DEIMOS instrument on the Keck~II telescope for 1 hr each, 
using the 1200 line mm$^{-1}$ grating (which has a dispersion of 0.33~\AA\
pixel$^{-1}$) centered at 7800~\AA.  This led to an average spectral coverage 
of 6450\,--\,9150~\AA.  

Modified versions of the {\tt spec2d} and {\tt spec1d} software\footnote{http://astron.berkeley.edu/$\sim$cooper/deep/spec2d/primer.html,
\newline\indent
http://astron.berkeley.edu/$\sim$cooper/deep/spec1d/primer.html
} developed by the DEEP2 team
at the University of California, Berkeley were used to reduce and analyze the
spectroscopic data \citep[see][]{simon2007,gilbert2007}.  Reduced one-dimensional spectra are cross-correlated with
a library of template stellar spectra to determine the redshift of the object. 
Each spectrum is visually inspected and assigned a quality code based on the 
number and quality of the absorption lines.  A spectrum must have at least two 
identifiable spectral features to be considered to have a secure 
redshift measurement \citep{guhathakurta2006,gilbert2007}.  A 
heliocentric correction and a correction for possible miscentering of the star 
within the slit (based on the position of the telluric A-band relative to 
atmospheric emission lines in the observed spectrum) is applied to the 
measured velocities \citep{simon2007,sohn2007}. 
The median velocity error is $\sim 6$~\kms, based on error estimates from 
the cross-correlation routine and repeat measurements of individual stars. 

\subsection{Selection of M31 RGB Stars}
 
The line-of-sight velocity distributions of foreground MW dwarf stars
and M31 RGB halo stars overlap, making it nontrivial to identify 
individual stars as M31 red giants or MW dwarfs. We use the diagnostic 
method described in \citet{gilbert2006} to isolate M31 RGB stars
from foreground MW dwarf stars in our spectroscopic sample.  Empirical 
probability distribution functions in five photometric and spectroscopic 
diagnostics are used to determine the likelihood an
individual star is an M31 red giant.  These diagnostics include 
(1) line-of-sight velocity, 
(2) photometry in the $M$, $T_2$, and DDO51 bands (when available), 
(3) the EW of the \nai\ doublet at 8190~\AA\ versus the \vio\ color of the star, 
(4) position of the star in an \ivi\ color magnitude diagram, and 
(5) photometric versus spectroscopic metallicity estimates. 
The M31 RGB sample used in this analysis includes all stars that are 
more probable to be M31 red giants than MW dwarfs.

\subsection{Metallicity Estimates}\label{sec:data_met}
Metallicity estimates (\feh) are derived by comparing a star's position in the \ivi\ 
color magnitude diagram with theoretical isochrones adjusted
to the distance of M31 \citep[783~kpc;][]{stanek1998,holland1998}.  Interpolation within the 
grid of 12~Gyr, [$\alpha$/Fe]=0, \citet{vandenberg2006} isochrones yields an \feh\ measurement
for each M31 RGB star \citep{kalirai2006halo}.  

Spectroscopic estimates of the metallicity
of an individual star can also be obtained using the equivalent width of the \caii\
triplet and the calibration relations of \citet{rutledge1997pasp1,rutledge1997pasp2}.  On average, the 
two estimates are in good agreement for the M31 survey data \citep{kalirai2006halo}, although there is 
significantly larger scatter in the spectroscopic metallicity estimates 
due to large measurement errors in the \caii\ EW measurements 
\citep{kalirai2006halo,gilbert2006}.  This is due primarily to the relatively 
low average signal-to-noise ratio of M31 RGB spectra in our spectroscopic 
survey (typical values of $S/N\sim 10$~pixel$^{-1}$).  There is 
smaller scatter in the spectroscopic metallicity estimates and very good 
agreement with photometric metallicity estimates for our highest $S/N$ M31 
RGB spectra.  Due to the large measurement 
uncertainties in the spectroscopic
metallicity estimates for individual stars, we choose to use photometric 
metallicities for this analysis.  

The use of photometric metallicities introduces systematic biases due to the isochrone set used and to 
the adoption of a single age and alpha-enrichment for the stellar populations. 
Estimates of \feh\ based on different sets of isochrones vary by less than 0.15~dex \citep{kalirai2006halo}.
Although the alpha-enrichment of M31's stellar halo has yet to be 
observationally constrained, deep HST/ACS observations of fields in M31's stellar halo which
reach the main sequence turnoff \citep[e.g.,][]{brown2006apj,brown2007} have shown 
that the stellar population of M31's halo spans a wide range in age.  
In the SE shelf and H13s fields, which are contaminated by GSS debris, the stellar population ranges from 6\,--\,13~Gyr in age.   
The younger stars are found to be the most metal-rich, while the 
old population is relatively metal-poor \citep{brown2006apjl,brown2006apj}.  In an HST/ACS M31 halo field at 21~kpc, a spread in ages of $\sim 8$\,--\,13~Gyr is found.  Once again, the younger stars are the most metal-rich while the old stellar population is relatively metal-poor.  
We therefore expect our metallicity estimates to be affected mostly at the metal-rich end by the assumption of a single, old stellar population.  Varying the age from 12 to 6~Gyr results in an $\approx +0.15$~dex
shift in the \feh\ estimates derived from isochrone fitting for metal-rich (\feh$>-1$) stars \citep{gilbert2009gss}.   

\subsection{Identification of Substructure: Kinematics and Metallicity}\label{sec:data_ident}

\begin{figure}
\epsscale{1.0}
\plotone{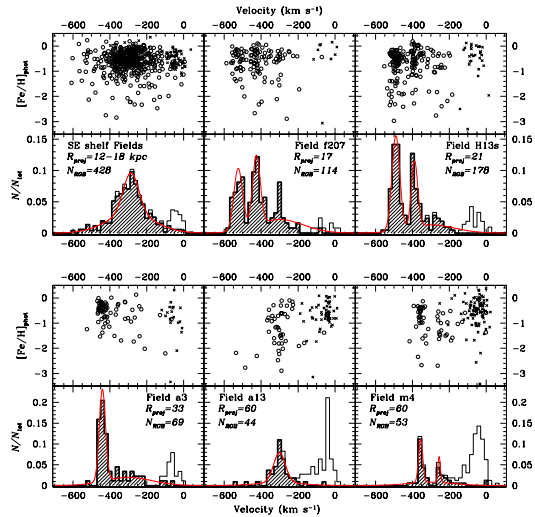}
\caption{Line-of-sight velocity distributions and \feh\ versus line-of-sight 
velocity for the fields discussed in this paper.  Shaded velocity 
histograms and open circles represent stars classified as M31 red giants 
by the \citet{gilbert2006} diagnostic method, while crosses represent 
stars classified as MW dwarfs. Open velocity histograms show the 
velocity distribution of all the stars (M31 RGB and MW dwarf) in our 
fields.  The solid curves show the best-fit velocity distributions in each field (\S\,\ref{sec:data_ident}).  
In general, the observed kinematically cold populations are more metal-rich 
than the rest of the stellar population.  }
\label{fig:data}
\end{figure}

Figure~\ref{fig:data} presents line-of-sight velocity distributions and 
\feh\ versus line-of-sight velocity for the stars in each of six fields 
in which our survey of M31's stellar halo has discovered kinematical substructure.  The distributions of CMD-based \feh\ values (\S\,\ref{sec:data_met}) 
in Figure~\ref{fig:data} show the presence of a relatively metal-poor 
(\feh$\lesssim -1$) population with a broad velocity distribution in 
each field, while the kinematically cold substructure tends to be relatively 
metal-rich (\feh$> -1$). 
Detailed analyses of the kinematics, chemical 
properties, spatial distributions, and plausible origins of the various substructures are presented  
elsewhere [SE shelf region in \citet{gilbert2007}; a3 in \citet{guhathakurta2006}; 
H13s in \citet{kalirai2006gss}; fields f207, H13s, a3, a13, and m4 in \citet{gilbert2009gss}].  

The primary signature used to identify substructure in a field is the 
presence of a kinematically 
cold population; the line-of-sight velocity distributions of these 
fields are inconsistent with being drawn from a single, broad 
Gaussian centered at M31's systemic velocity \citep[\mv=$-300$~\kms, 
$\sigma_v=129$~\kms;][]{gilbert2007}.
Fields a13 and m4
are at larger projected radial distances from the center of M31 ($R_{\rm proj}\sim60$~kpc) than
the study of \citet{gilbert2007}.  Their 
line-of-sight velocity distributions are also inconsistent with being drawn 
from a Gaussian with parameters \mv=$-300$~\kms, $\sigma_v=100$~\kms, the velocity dispersion implied for the distance of these fields by the study of \citet{chapman2006}. 

The velocity distribution in each field has been fit by a combination 
of Gaussians using a maximum-likelihood technique (red curves in 
Figure~\ref{fig:data}).  In general, the number of Gaussians used in the 
multi-Gaussian fits is the minimum required to produce a maximum-likelihood 
fit that is consistent with the data, although 
additional evidence of substructure in a field is also taken into account in
determining the number of kinematical components in a given spectroscopic field \citep{gilbert2007,gilbert2009gss}.
The Gaussian components of the underlying kinematically
hot population were held fixed (\mvsph\,$=-300$~\kms, \sigvsph\,$=129$~\kms)  
while the properties of the 
Gaussians representing the kinematically cold populations (mean velocity and
velocity dispersion) and the fraction
of stars in each cold population were allowed to vary.  
The SE shelf, f207, H13s, a3, and a13 fields all contain 
substructure related to the GSS in M31 
\citep{kalirai2006gss,guhathakurta2006,gilbert2007,gilbert2009gss}.  
Fields f207 and H13s both have
 a second kinematically cold component which may be unrelated to 
the GSS.  The cold components with mean velocities of $-527$~\kms\ and $-490$~\kms\ 
in f207 and H13s, respectively, are those associated with the GSS \citep{gilbert2009gss}. 
Although there are some lines of evidence that 
suggest the substructure in m4 and the secondary substructures in fields f207 and H13s 
may also be associated with the GSS \citep{fardal2008,gilbert2009gss}, the origins 
of these features are currently unknown.


\begin{table}[tb!]
\begin{center}
\caption{Fraction of M31 RGB Stars in Kinematically Cold Components.}
\vskip 0.3cm
\begin{tabular}{lll}
\hline
\hline
\multicolumn{1}{c}{Field} & \multicolumn{1}{c}{\mvsb} & \multicolumn{1}{c}{$N_{\rm KCC}/N_{\rm tot}$\tablenotemark{a}}\\
& \multicolumn{1}{c}{(\kms)} & \\
\hline
SE shelf & $-285$ & $0.41^{+0.10}_{-0.09}$ \\
f207 & $-524$ & $0.31^{+0.08}_{-0.07}$ \\
f207 & $-426$ & $0.31^{+0.11}_{-0.10}$ \\
H13s & $-490$ & $0.48^{+0.07}_{-0.06}$ \\
H13s & $-388$ & $0.27^{+0.07}_{-0.08}$ \\
a3 &  $-441$ & $0.59^{+0.11}_{-0.12}$ \\
a13 & $-301$ & $0.72^{+0.15}_{-0.21}$ \\
m4 & $-355$ & $0.45^{+0.14}_{-0.13}$ \\
m4 & $-255$ & $0.16^{+0.10}_{-0.08}$ \\
\hline
\hline
\end{tabular}
\tablenotetext{a}{Quoted errors are the 90\% confidence limits from the maximum likelihood analysis (\S\,\ref{sec:data_ident}).}
\label{table2}
\end{center}
\end{table}

\subsection{Characteristics of Observed Substructure: Surface Brightness versus Metallicity}\label{sec:char_obs_sub}
\begin{figure}
\plotone{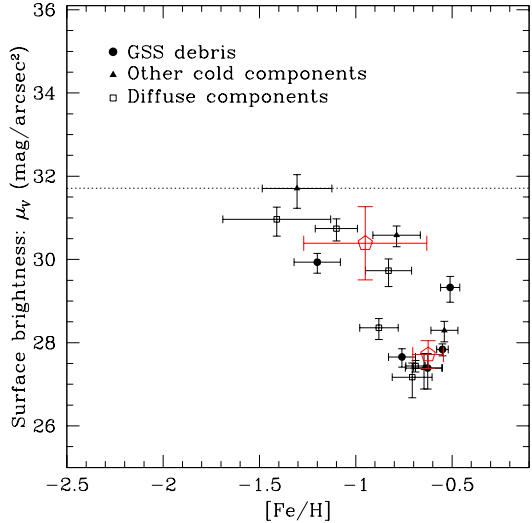}
\caption{Surface brightness versus average metallicity for each of the 
kinematical components in Figure~\ref{fig:data}: kinematically cold components likely associated 
with the GSS (solid circles), kinematically cold components which are not known to be associated 
with the GSS (solid triangles), and the kinematically hot components (open squares).
The error bars represent the Poisson error for the surface brightness 
estimates and the error in the mean \feh\ value.  
The mean and root-mean-square (rms) [Fe/H] and $\mu_{V}$ of the lower surface brightness ($\mu_{V}>29$~mag arcsec$^{-2}$) and higher surface brightness ($\mu_{V}<29$~mag arcsec$^{-2}$) kinematically cold components are also shown (open hexagons and error bars).
Higher surface brightness features are 
on average more metal-rich than lower surface brightness features, and kinematically cold substructures are on average more metal-rich than the kinematically hot populations.  Lower surface brightness features display a larger spread in metallicity than higher surface brightness features.  The dotted line denotes the faintest identifiable tidal streams to date, given the number of stars in our DEIMOS survey.  }
\label{fig:mufeh_data1}
\end{figure}

Figure~\ref{fig:mufeh_data1} displays surface brightness versus mean 
metallicity for the M31 
stellar halo data presented in Figure~\ref{fig:data}.  Each kinematical
component identified in the 6 fields shown in Figure~\ref{fig:data} 
is represented: the kinematically cold
components related to the GSS (solid circles), 
the kinematically cold components that are not known to be associated with the 
GSS (solid triangles), and the kinematically
hot population (open squares).  

Table~\ref{table2} lists the maximum-likelihood 
fraction of M31 RGB stars in each of the kinematically cold components 
(from the fits shown in Fig.~\ref{fig:data}).  These
fractions were used to determine the number of M31 RGB stars in each component.  
The ratio
of M31 RGB to MW dwarf stars, multiplied by the expected number of MW
dwarf stars along the line of sight to each field \citep[based on the Besancon
Galactic population model;][]{robin2003} is used to estimate the surface brightness of each 
component, as described in \citet{guhathakurta2005}.  Correction factors that account for the rate of 
successful velocity measurements as a function of magnitude and 
varying M31 RGB target efficiency due to photometric pre-selection of stars likely to 
be red giants are applied to each field (Gilbert et al., in preparation).  
The spectroscopic surface brightness estimates are normalized using photometric surface brightness estimates 
from \citet{pritchet1994}.  Error bars
represent Poisson error estimates based on the number of stars in each component.    

The \feh\ for the kinematically cold components is the mean \feh\ 
of stars within $\pm2\sigma_v$ of the mean velocity of the cold component 
($|v_{\rm star}-$\,\mvsb$|<2$\,\sigvsb), while the \feh\ for the 
kinematically hot components is the mean \feh\ of stars with velocities 
greater than 2\sigvsb\ from the mean velocity of the cold components 
($|v_{\rm star}-$\,\mvsb$|>2$\,\sigvsb; in 
fields with two cold components, stars must meet this criterion for 
both).  The numbers of stars which are thus used for the \feh\ 
measurement for each component are consistent with the number of stars
expected to be in each component based on the maximum-likelihood fits 
shown in Figure~\ref{fig:data}.  Error bars represent the error in the mean \feh\ value.

Since stars cannot be identified individually as belonging to a particular 
kinematical component, these velocity cuts are chosen to maximize the number of 
stars involved in the \feh\ measurement of the kinematically cold component, 
while minimizing contamination in the 
mean \feh\ measurement from other kinematical components.   
Even so, the metallicity estimate of the substructure
will be biased by contamination from the kinematically hot component.
This effect will be greatest for the least dominant substructures, and will in
general bias the mean metallicity of the kinematically cold substructures to lower
values.  In the SE shelf fields, where we have a large statistical sample, 
accounting for the contamination by the kinematically hot component 
in the measurement of the mean 
metallicity of the kinematically cold component changes the mean 
metallicity by only $+0.03$~dex \citep{gilbert2007}; however, it must be noted that
the \feh\ distribution of the kinematically hot population in the SE shelf fields
is very similar to the \feh\ distribution of the kinematically cold population. 

The assumption of a single, old stellar population likely provides a larger source 
of systematic error in the photometric (CMD-based) metallicity 
estimates.  As discussed in 
\S\,\ref{sec:data_met}, this assumption is known to be wrong for at 
least some of our fields and is expected to affect our metallicity estimates mostly on the 
metal-rich end.  Indeed, a comparison of the mean \caii-based spectroscopic metallicity 
estimates (measured in bins of photometric metallicity) of stars associated with the kinematically 
hot and cold populations shows that the kinematically cold populations do, on average,
have a higher spectroscopic abundance than the kinematically hot population at a  given
photometric abundance.  This indicates that the substructure in M31's stellar halo is
on average younger and/or more enriched in [$\alpha$/Fe] than M31's kinematically smooth stellar halo.
Using 6~Gyr instead of 12~Gyr isochrones changes the \feh\ estimates by $\sim +0.15$ for more metal-rich 
stars (\feh$\gtrsim-1.0$) and by as much as $\sim +0.3$ for metal-poor stars (\feh$\lesssim-1.5$).      
Thus, the true metallicity spread is likely greater than Figure~\ref{fig:mufeh_data1} suggests.       

The data in Figure~\ref{fig:mufeh_data1} show a clear correlation 
between surface brightness and metallicity: higher surface brightness 
features are more metal-rich than lower surface brightness features.  The 
kinematically cold substructures are on average more metal-rich 
than the underlying kinematically hot components, as was
implied in Figure~\ref{fig:data}.   Figure~\ref{fig:mufeh_data1} also shows
evidence for a larger spread in metallicities among lower surface brightness 
features ($\mu_{V}>29$~mag arcsec$^{-2}$) than higher surface brightness features ($\mu_{V}<29$~mag arsec$^{-2}$).  The root-mean-square (rms) 
deviation of average [Fe/H] values among higher surface brightness, kinematically cold 
components is only 0.08~dex, while the rms deviation of average [Fe/H] values among 
lower surface brightness, kinematically cold components is 0.32~dex.

\begin{figure}
\plotone{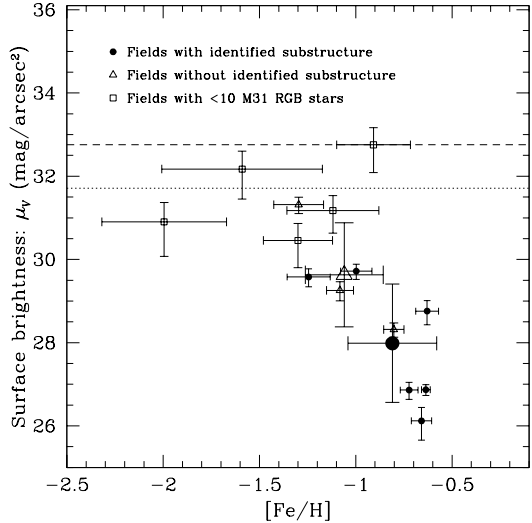}
\caption{The same as Figure~\ref{fig:mufeh_data1} but for entire fields from our M31 spectroscopic survey.  Fields are divided into three categories: fields 
with (solid circles; Fig.~\ref{fig:data}) and without (open triangles) 
kinematically identified substructure, and fields with small numbers 
of M31 RGB stars 
($\lesssim 10$), for which we cannot determine if substructure is present.  
The large symbols show the mean and rms (errorbars) [Fe/H] and $\mu_{V}$ of fields with and without identified substructure.  Fields with identified substructure are on average more metal-rich than fields without identified substructure.
The dashed
line represents the current surface brightness limit achieved in our 
M31 spectroscopic survey, while the dotted line denotes the faintest 
tidal stream identified in our survey (Fig.~\ref{fig:mufeh_data1}).   
}
\label{fig:mufeh_data2}
\end{figure}

Figure~\ref{fig:mufeh_data2} displays surface brightness
versus mean metallicity for fields from our M31 halo spectroscopic survey with 
kinematically-identified substructure (those presented in Figure~\ref{fig:data}; 
solid circles), fields 
without kinematically-identified substructure (open triangles), and fields for which 
we cannot determine if substructure is present due to small 
numbers of M31 RGB stars ($\lesssim 10$ M31 stars per field; open squares).   
The fields with small numbers of M31 RGB stars are all located in the outer
 regions of M31's stellar halo ($R_{\rm proj}\gtrsim 80$~kpc), while the fields without 
 kinematically-identified substructure span a similar range in projected distance 
 from M31's center as the fields with kinematically-identified substructure.
Fields with identified substructure are on average more metal-rich than 
fields without identified substructure, as would be expected if kinematically cold 
substructures are more metal-rich on average than the kinematically hot population 
(Fig.~\ref{fig:mufeh_data1}).

\section{Simulated Stellar Halos}\label{sec:sims}

Simulations of hierarchical stellar halo formation provide a physical explanation of the surface brightness versus metallicity trend observed in our M31 halo data.  In this section, we briefly describe the simulations (\S\,\ref{sec:sims_method}), present the abundance properties of substructure in the simulated stellar halos (\S\,\ref{sec:sims_properties}), and discuss the physical mechanisms behind the trends (\S\,\ref{sec:sims_interp}).  We will compare the abundance trends in the simulations with our M31 observations in \S\,\ref{sec:comparison}.

\subsection{Methods}\label{sec:sims_method}
The BJ05 simulations follow the formation of 11 MW-analog stellar halos 
via accretion in a cosmological context.  
For each host halo, a merger tree was generated using the extended
Press-Schechter formalism \citep{somerville1999,lacey1993}.  
Only merger histories without recent high-mass mergers were selected, resulting
in $z=0$ halos suitable for hosting a large disk galaxy like the Milky Way or M31.
For each merger, an $N$-body simulation was used to track the evolution 
of the dark matter halo of the satellite galaxy in the
host halo potential.  The potential of the host galaxy is modeled with
a smoothly growing, analytic function, and satellite-satellite interactions
are neglected.  Overall, these assumptions will lead to an unnatural enhancement of the  level of substructure in the simulated halos, especially in the inner halo.  
Thus, in the
following analysis we discuss only the properties of the simulated stellar 
halos exterior to $R=20$~kpc. 

Gas accretion onto each satellite halo tracks its mass
accretion history and the star formation rate within each object is 
proportional to its instantaneous gas content over a fixed timescale.
The timescale is chosen
so that dwarfs infalling today would contain similar gas fractions to those observed in
Local Group field objects (see BJ05 for details).  
Given the gas infall, star formation rates, and mass of a satellite as a function of time, the method of
\citet{robertson2005} is used to follow the chemical evolution of the stellar
populations \citep[see also][]{font2006apjhalos}.
The \citet{robertson2005} prescriptions
include enrichment from Type Ia and Type II supernovae and the effects of feedback
(from supernovae and stellar winds).  
The free parameters in the prescriptions are
constrained by requiring consistency with observations: the Fe and $\alpha$-element wind
efficiencies are chosen separately by requiring matches to the observed stellar
mass-metallicity relation for dwarf galaxies \citep[e.g.,][]{larson1974,mateo1998,dekel2003} 
and the $\alpha-$element abundance patterns in satellite dwarf spheroidals \citep[e.g.][]{venn2004}.
Both gas accretion and star formation are
assumed to truncate once the satellite is accreted onto the host halo.
A variable mass-to-light ratio is assigned to every dark
matter particle in order to embed the stellar components within the
dark matter halos of the satellites in a way that reproduces the observed distribution of structural properties of Local Group dwarfs.
Each stellar particle in the simulations is randomly assigned \feh\ and 
[$\alpha$/Fe] ratios based on the satellite's star
formation history \citep{font2006apjhalos}.  Thus each model 
satellite contains a realistic intrinsic metallicity spread, but no radial
metallicity gradient.

The resulting simulated halos have similar luminosities and stellar density
profiles to the Milky Way's halo, and the surviving satellites are
comparable in number, luminosity distribution, and structural properties to those of the Milky Way (discovered prior to 2005).  The models are also able
to reproduce the observed difference in [$\alpha$/Fe] ratios between the
Milky Way's stellar halo and dwarf satellites. The lower [$\alpha$/Fe]
ratios in the surviving simulated dwarf galaxies are due to their late accretion times
and the prolonged, less bursty star formation histories of these systems
compared with those of the progenitors of the halo \citep{robertson2005,font2006apjsats}. In the models, the kinematically cold
streams tend to have distinct chemical abundances (i.e. higher [Fe/H] and
lower [$\alpha$/Fe]) than that of the smooth, more well-mixed halo \citep{font2006apjhalos}. This is also due to the fact that the progenitors of these streams were accreted following the formation of the bulk of the halo.

Of course, in the framework of the hierarchical formation of stellar halos,
there is no clear demarcation between tidal streams and the ``smooth''
halo. In reality, present day stellar halos are expected to display a
natural progession in the kinematical properties of their unbound
substructure, ranging from thin and kinematically cold to more
disperse, kinematically hot substructure. In light of the above findings,
the prospect of using chemical abundances as a complementary dimension to
phase-space to disentangle various substructures is encouraging. 

\subsection{Predicted Abundance Properties of Substructure}\label{sec:sims_properties}
 
\begin{figure}
\begin{center}
\plottwo{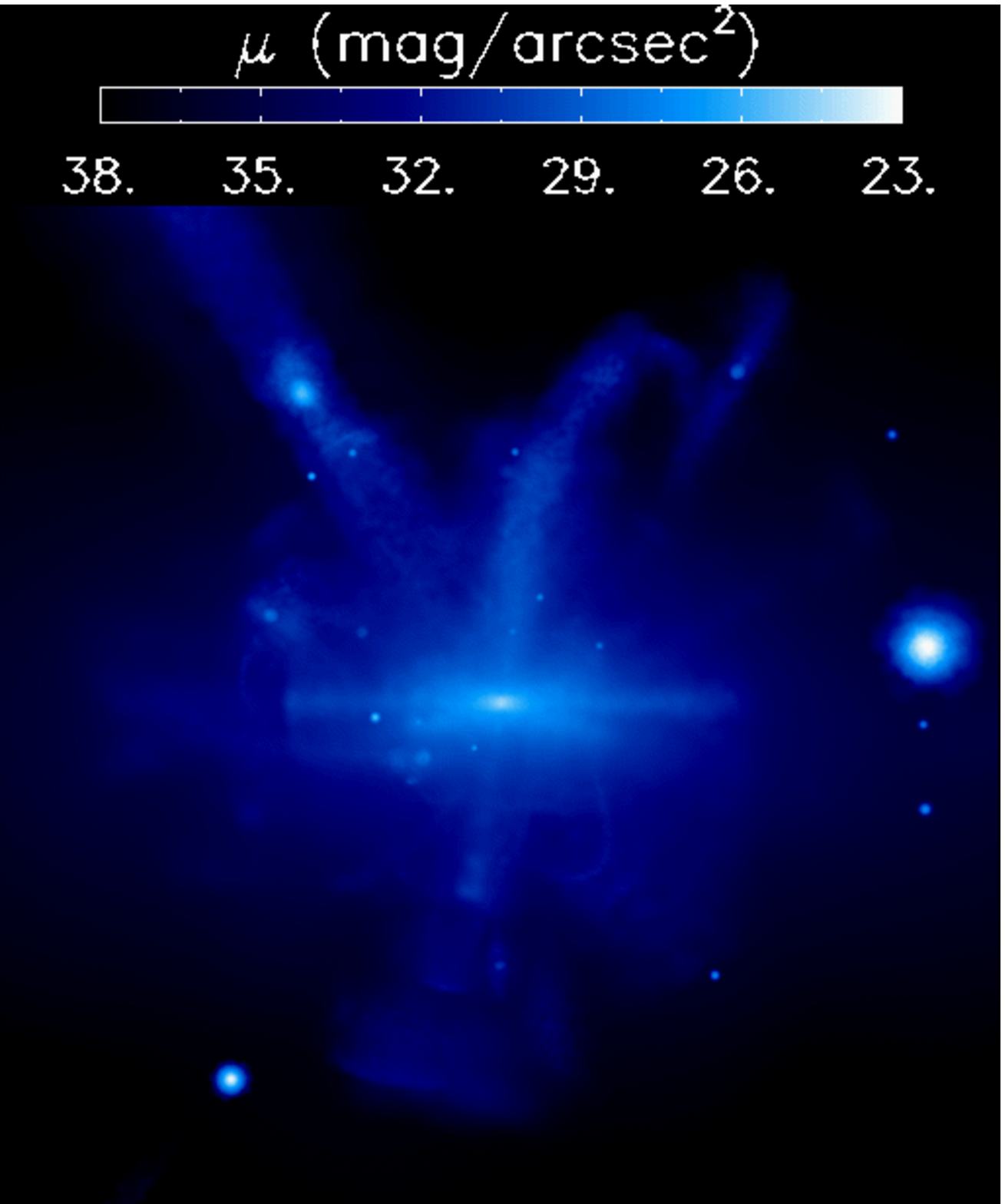}{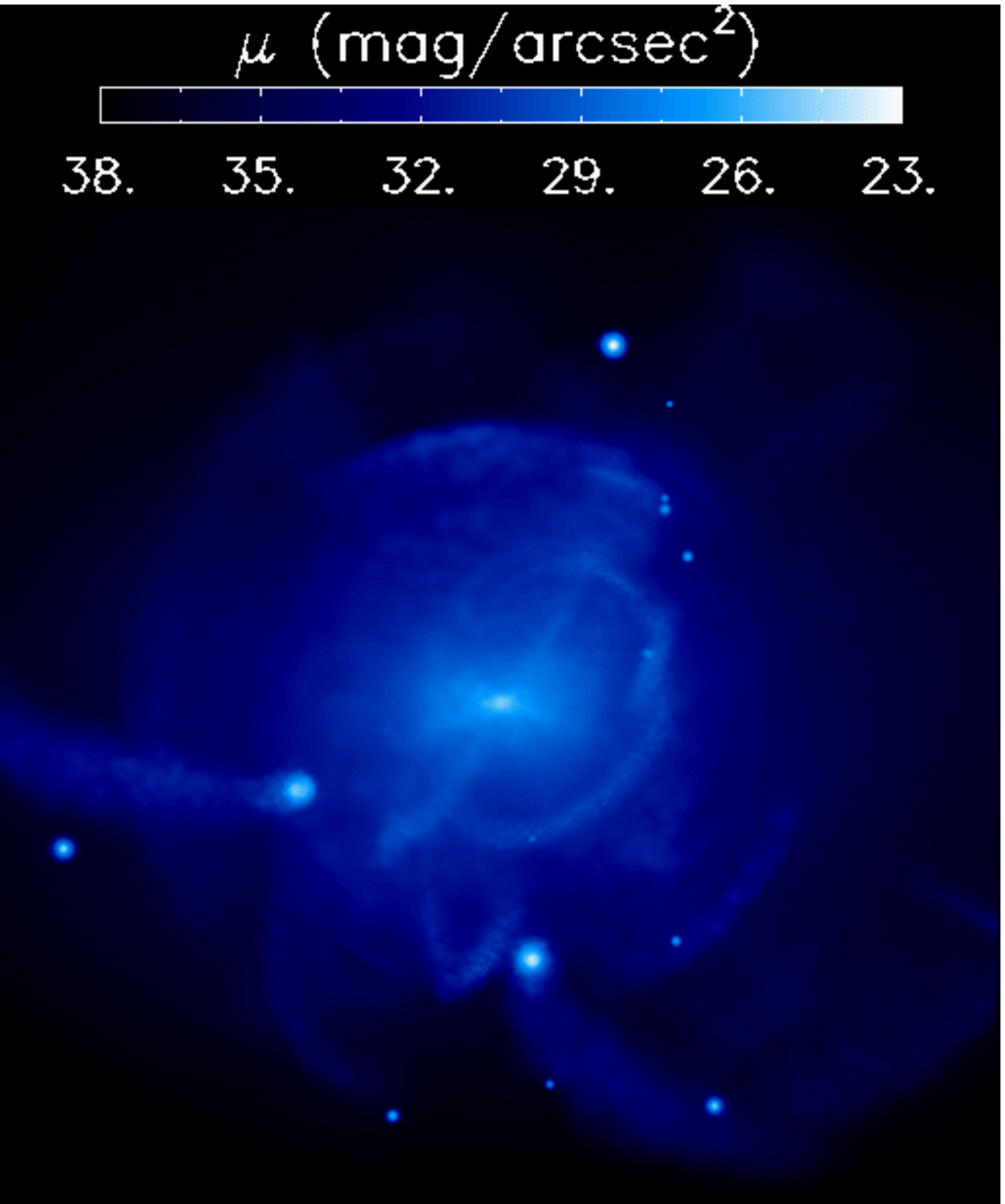}\\
\plottwo{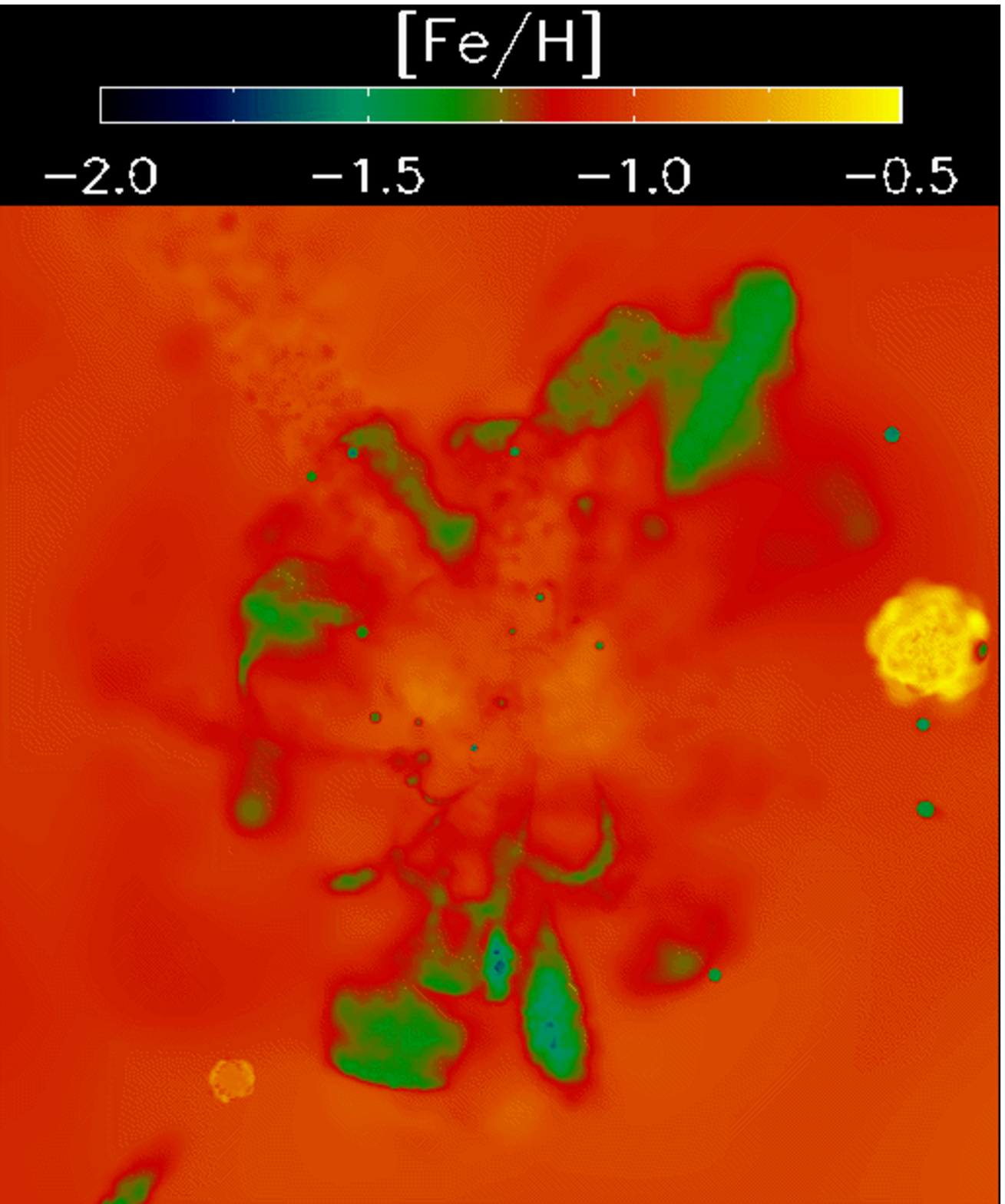}{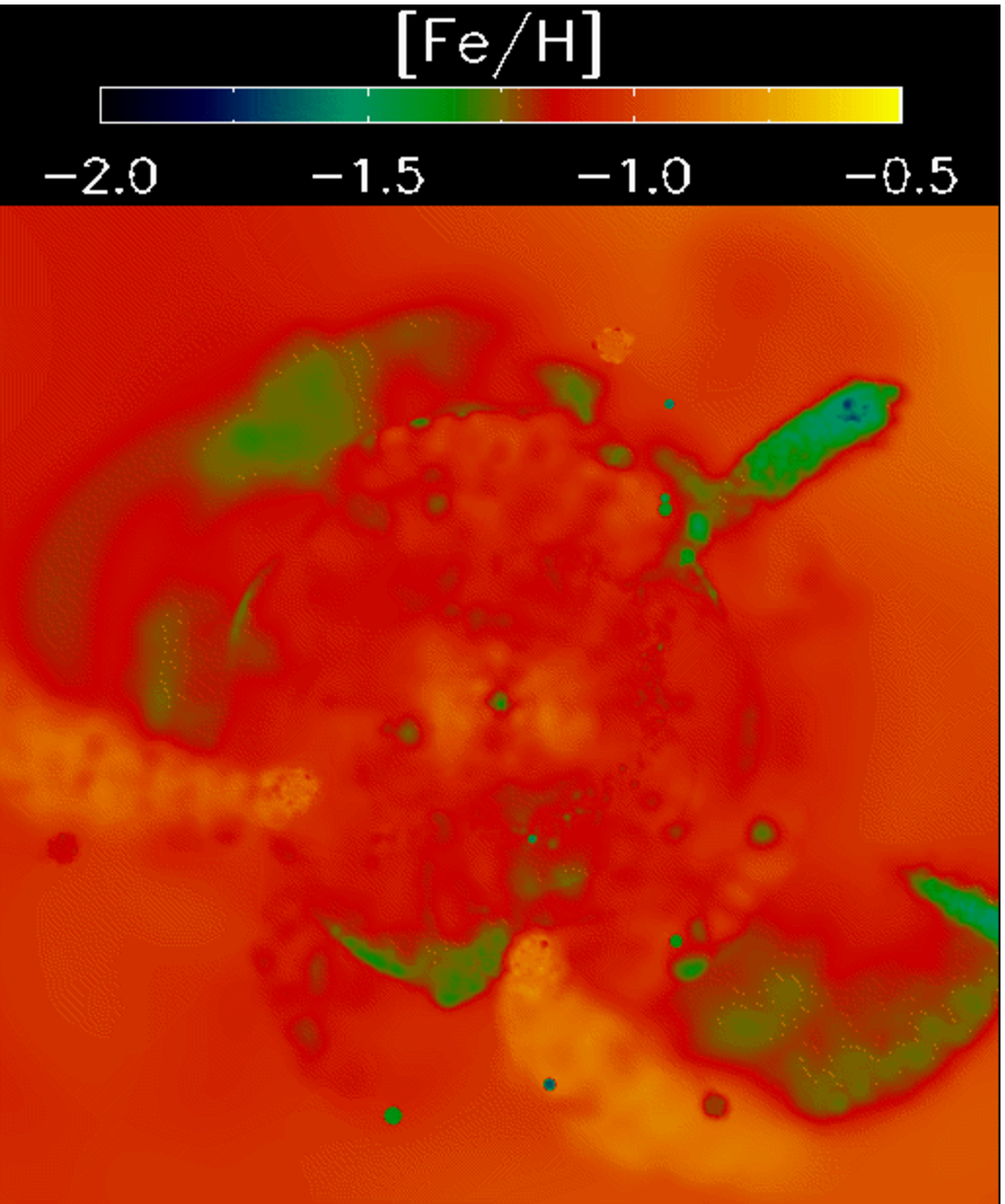}\\
\plottwo{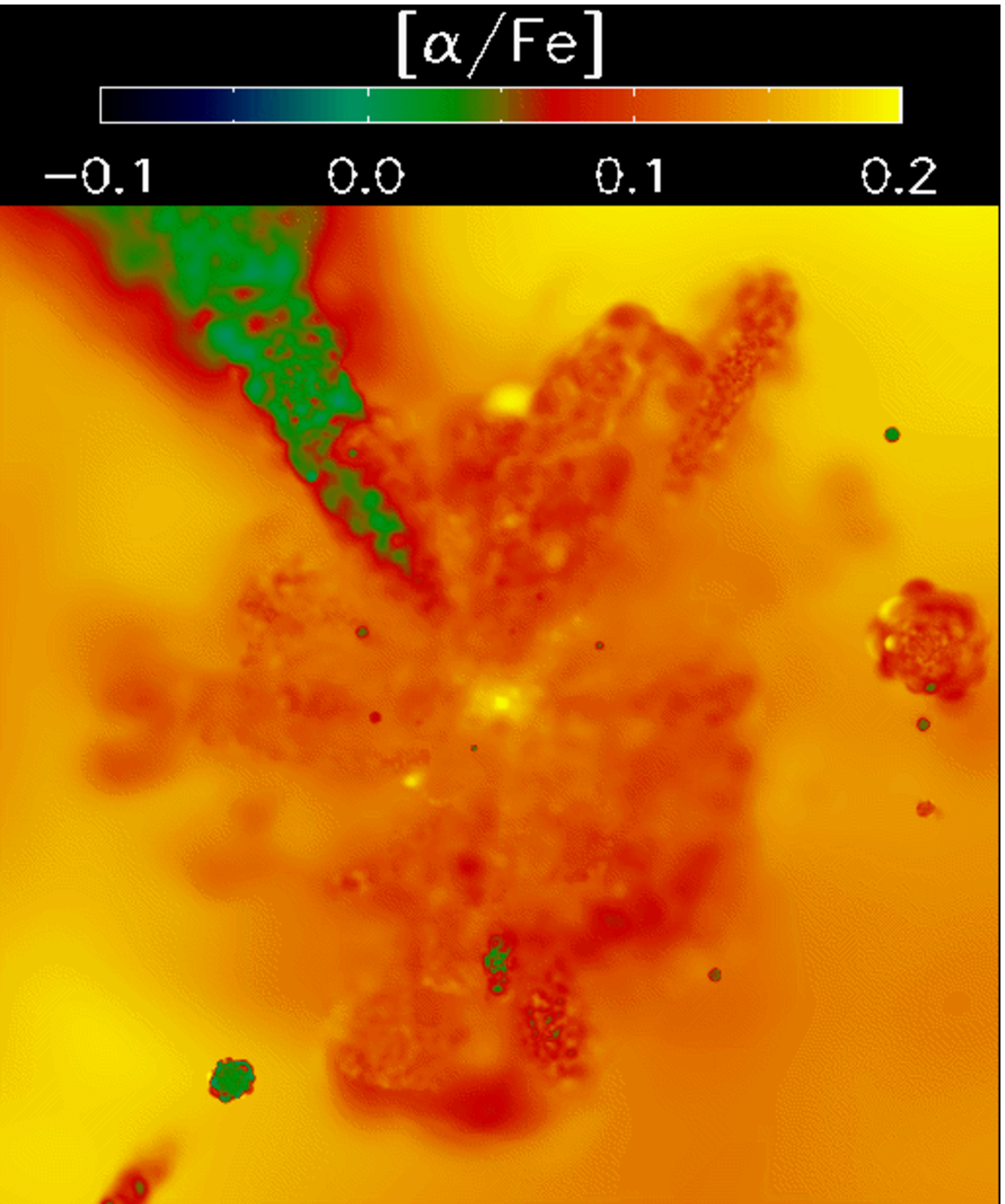}{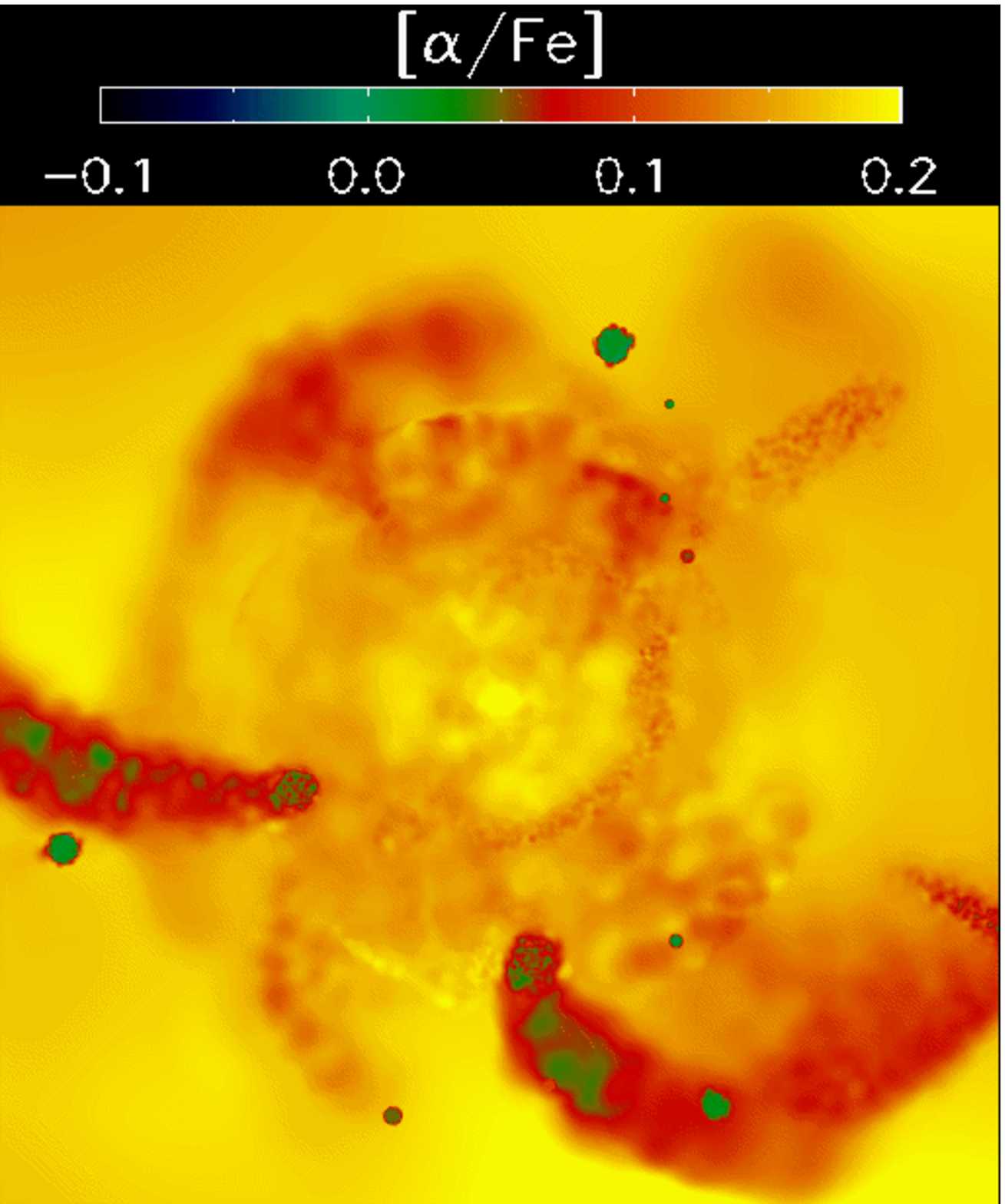}
\end{center}
\caption{
Maps of projected surface brightness ({\it top}), 
average metallicity ({\it middle}), 
and average alpha-enrichment ({\it bottom}) 
for 2 of the BJ05 simulated stellar halos.  Each map depicts a 300$\times$300~kpc region. 
The \feh\ and [$\alpha$/Fe] values are the stellar-mass-weighted average of 
all the stars in each pixel.  A comparison of the surface brightness 
and metallicity maps reveals that the brightest features tend to be 
more metal-rich and less alpha-enhanced than the rest of the halo. Image credit: Sanjib Sharma. 
}
\label{fig:sims}
\end{figure}

Figure~\ref{fig:sims} shows the projected surface brightness 
({\it top}), metallicities ({\it middle}), and alpha-enrichment ({\it bottom})  for two of the BJ05 
simulated stellar halos.  
The projected surface brightness reflects the total luminosity in each 
0.25~kpc~$\times$~0.25~kpc 
pixel, summed over all the stellar populations.  The \feh\ and [$\alpha$/Fe] values 
are the stellar-mass-weighted average of the stellar populations in each pixel.  These maps reveal that 
the highest surface brightness streams tend to be more metal-rich and less alpha-enhanced than the lower surface brightness streams.  

\begin{figure}
\epsscale{1.0}
\plotone{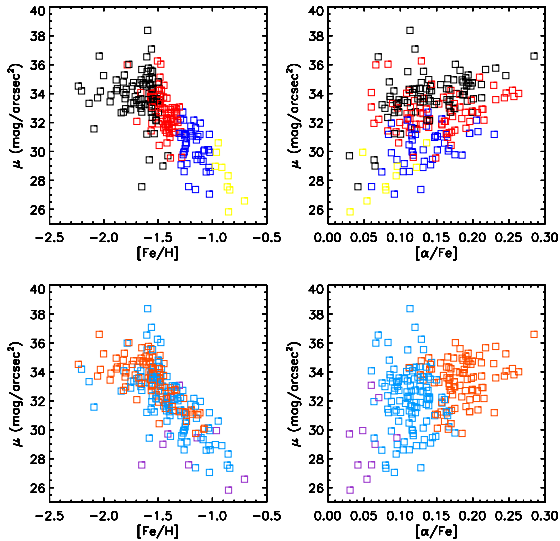}
\caption{Maximum surface brightness versus average metallicity ({\it left panels})
and alpha enhancement ({\it right panels}) for stellar streams in the 11 BJO5 
simulated stellar halos. The top panels are color coded according to the luminosity
of the progenitor satellite [$L<10^6$\Lsun\ ({\it black}); $10^6$\Lsun\,$<L<10^7$\Lsun\ 
({\it red}); $10^7$\Lsun\,$<L<10^8$\Lsun\ ({\it blue}); $10^8$\Lsun\,$<L<10^9$\Lsun\ ({\it yellow})]
while the bottom panels are color coded according to the time of accretion 
of the progenitor satellite 
onto the host halo [ $3<t_{\rm acc}<6$ Gyr 
({\it purple}); $6<t_{\rm acc}<9$ Gyr ({\it cyan}); $9<t_{\rm acc}<12$~Gyr 
({\it orange})].   The distribution in the left panels is 
a reflection of the mass-metallicity relationship observed for 
Local Group dwarfs, while the  
distribution in the right panels reflects the
time-sensitivity of alpha elements.  The bottom panels also show a 
slight tendency for more recent accretion events to be brighter. 
}
\label{fig:mfsim}
\end{figure}

This is demonstrated quantitatively in Figure~\ref{fig:mfsim}, 
which displays maximum 
surface brightness versus mean metallicity ({\it left panels}) and alpha-enhancement ({\it right panels}) for 
individual stellar streams produced by accretion events in all 11 BJ05
stellar halos, taken from the analysis of \citet{johnston2008}. 
Only 
stellar streams that contribute at least 50\% of the total light in the halo
at the position of their highest surface brightness point and which are at 
radial distances greater than 30~kpc from the center of the galaxy are 
included.  Although for any given maximum surface brightness stellar streams exhibit a range of mean \feh\ and [$\alpha$/Fe] values, the higher surface brightness streams are in general more metal-rich and less alpha-enhanced than the lower surface brightness streams.

\subsection{Physical Interpretation}\label{sec:sims_interp}

The points in Figure~\ref{fig:mfsim} are 
color-coded by mass of the progenitor satellite ({\it top panels}) and time
of accretion of the progenitor satellite onto the host halo ({\it bottom panels}; see figure caption for details).  The top left panel clearly shows the effect of the
assumption of a mass-metallicity relation for the progenitor satellites (based on observations of dwarf galaxies; \S\,\ref{sec:sims_method}): more massive satellites are more metal-rich, thus the tidal streams produced as massive satellites fall into the host halo are also more metal-rich.  Since the more massive progenitors also typically produce brighter stellar streams, it follows that the surface brightness and metallicity of stellar streams are correlated. 

The alpha-enhancement of the stellar streams traces the time of accretion of the progenitor onto the host halo, as is
demonstrated in the bottom right panel of Figure~\ref{fig:mfsim}. 
Alpha elements are produced in Type II supernovae; their abundances become subsequently 
diluted by the elements produced in Type I supernovae.  Thus, progenitors accreted
early are alpha-enriched relative to progenitors that are accreted later and have had longer star formation histories.  
The surface brightness of stellar streams observed today is a function 
not only of the mass of the progenitor but also of the time since accretion, 
therefore the bottom right panel shows a slight trend for stellar streams produced by progenitors that were accreted at early times to have lower 
surface brightnesses than progenitors that were accreted at later times.  This leads to the anti-correlation between surface brightness and alpha-enrichment.  

\section{Comparison of Observations and Simulations}\label{sec:comparison}

\subsection{Observing the Simulations}\label{sec:comp_obssims}
Figure~\ref{fig:mfsim} provides useful insight into the physical mechanisms underlying the expected general relationships
between the chemical enrichment and maximum surface brightness of stellar streams.  The data in 
Figure~\ref{fig:mufeh_data1} appear to agree qualitatively 
with the characteristics seen in the left panels of 
Figure~\ref{fig:mfsim}; identifiable (and thus fairly dominant and 
high surface brightness) substructures 
tend to be relatively metal-rich.  However, our Keck/DEIMOS spectroscopic 
masks provide information on the characteristics
of the population in isolated lines-of-sight through M31's stellar halo, 
rather than global information about individual streams. 
The data cannot be compared directly with Figure~\ref{fig:mfsim},
since the maximum surface brightness of an observed stellar stream can only
be determined by tracing the full extent of the stream in space.
The only stellar stream for which this measurement could
currently be made in M31 is the GSS.   

\begin{figure}
\plotone{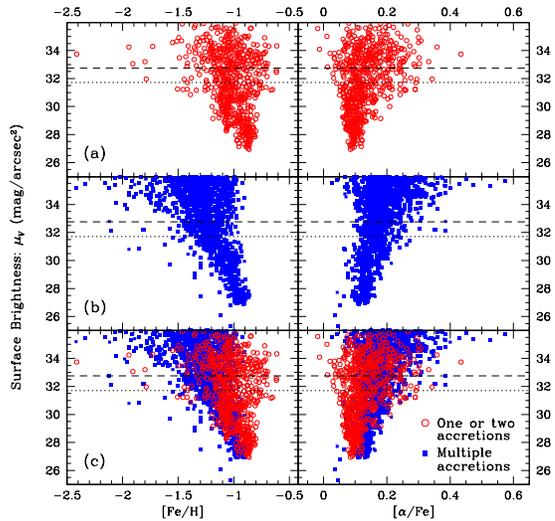}
\caption{Surface brightness versus stellar-mass-weighted average metallicity ({\it left}) and 
alpha-enhancement ({\it right}) for 
individual lines-of-sight through one of the 11 BJ05 simulated stellar halos,  
divided into two categories based on the
number and prominence of individual stellar streams.  ({\it a}) Lines-of-sight where the majority of the stellar population derives from one or two accretion events (solid red circles).  ({\it b}) Lines-of-sight where the majority of the stellar population derives from multiple ($\ge3$) accretion events  (open blue squares).  ({\it c}) The two types of sight-lines superimposed.  
The overall trends of surface brightness with metallicity and alpha-enhancement (panel {\it c}), 
are similar to those in Figure~\ref{fig:mfsim}, although with a somewhat steeper slope.
For a given surface brightness, lines-of-sight whose stellar populations are dominated by debris 
from one or two 
accretion events are on average more metal-rich 
and less alpha-enhanced than lines-of-sight whose stellar populations are 
composed primarily by debris from multiple 
accretion events.  The surface 
brightness of the faintest tidal debris that has 
been identified as kinematically cold substructure in our Keck/DEIMOS survey 
is denoted by the dotted line (Fig.~\ref{fig:mufeh_data1}) while the 
current surface brightness limit of our spectroscopic survey is denoted by the 
dashed line (Fig.~\ref{fig:mufeh_data2}).  
}
\label{fig:sim_feh_alpha}
\end{figure}

A more direct comparison between the data and simulations can be made
by analyzing stellar populations along individual lines-of-sight through 
the simulated stellar halos. Figure ~\ref{fig:sim_feh_alpha} displays 
mean stellar metallicity and alpha-enhancement as a function of 
surface brightness for sight lines through one of the BJ05 simulated 
stellar halos; each point represents a line-of-sight that covers a 
projected area of 0.4~kpc~$\times$~0.4~kpc.  
Only lines of sight beyond 20~kpc in projection from the center of the galaxy are included (\S\,\ref{sec:sims_method}).  
Lines-of-sight through the simulated stellar halos whose stellar populations are dominated by debris from one or two
accretion events are denoted by red solid circles in Figure~\ref{fig:sim_feh_alpha}, while lines-of-sight whose stellar populations are largely comprised of debris from multiple ($\ge3$) 
accretion events are denoted by blue open squares.

The same general trends seen in individual streams (Fig.~\ref{fig:mfsim}) are repeated 
here, with a steeper slope: higher surface brightness lines-of-sight are on average more 
metal-rich and less alpha-enhanced than lines-of-sight of lower surface 
brightness.  The same physical mechanisms discussed in \S\,\ref{sec:sims_interp} are relevant here, however 
each point in Figure~\ref{fig:sim_feh_alpha} is averaging over stellar 
populations from {\it multiple} satellite progenitors.   
Thus, the trends in metallicity 
and alpha-enhancement with surface brightness are 
not as strong.

\subsection{Common Trends}\label{sec:comp_trends}

We can't know a~priori the number of streams that are along any 
particular observed sight-line in M31's stellar halo. 
However, separating the kinematically cold and hot components in fields
with identifiable substructure, as is done in Figure~\ref{fig:mufeh_data1}, 
effectively separates a single field into the equivalents of lines-of-sight dominated by stellar populations arising from a single accretion event
(kinematically cold components) and lines of sight dominated by stellar populations arising from many accretion events (kinematically
hot components).   In a stellar halo built in a purely dissipationless
fashion, a population with a large spread in velocities is most likely to be 
formed from the superposition of debris originating in multiple satellites 
that merged early with the host galaxy.  
The dotted line in Figure~\ref{fig:sim_feh_alpha} denotes the surface 
brightness of the faintest tidal debris that is  
identifiable as kinematically cold substructure in our Keck/DEIMOS survey 
(Fig.~\ref{fig:mufeh_data1}). 

The simulations (Fig.~\ref{fig:sim_feh_alpha}) show the same general trends seen in 
our M31 data (Fig.~\ref{fig:mufeh_data1}): (1) Higher surface brightness features are on average more metal-rich, and (2) at a given surface brightness 
lines-of-sight dominated by debris from one or two accretion events tend to be 
more metal-rich than those dominated by debris from multiple accretion events.  

We have discussed the first trend in \S\S\,\ref{sec:sims_properties}\,--\,\ref{sec:sims_interp}, and shown that in a stellar halo formed via accretion, it can be explained via the mass-metallicity relationship for dwarf galaxies: larger galaxies are more metal-rich and also produce the brightest tidal streams.  

This same physical mechanism can also explain the second trend. In order 
for tidal debris from a single satellite to dominate the stellar population 
in a line-of-sight through a halo, the tidal debris must be part of a particularly
bright stellar stream relative to the average surface brightness at 
its location in the halo.  It thus must originate from a relatively massive 
satellite (relatively metal-rich) and/or recent accretion event (low relative alpha-enhancement) compared to the accretion events
in a line-of-sight of equivalent surface brightness but composed primarily 
of debris from multiple progenitors.      

With multiple physical mechanisms (e.g., the size and orbit of the progenitor and the time since its accretion onto the host halo) affecting the surface brightness of individual streams, a significant spread in the chemical properties of individual sight-lines of a given surface brightness, as seen in Figure~\ref{fig:sim_feh_alpha}, is to be expected.  This is particularly relevant for faint lines-of-sight, since faint stellar streams can be formed by the recent accretion of a low mass progenitor or early accretion events whose debris streams have spatially dissipated, or both.  The brightest lines of sight, however, must be composed of debris from very massive and/or recent events, resulting in a narrower range in chemical properties.  

The data also show evidence of a narrower spread in metallicity among higher surface brightness streams/lines-of-sight and a larger spread in metallicity among fainter streams/lines-of-sight (Figs.~\ref{fig:mufeh_data1} \& \ref{fig:mufeh_data2}, \S\,\ref{sec:char_obs_sub}).  The rms deviation of mean [Fe/H] values for the high surface brightness ($\mu_{V}<29$~mag arcsec$^{-2}$) and low surface brightness ($\mu_{V}>29$~mag arcsec$^{-2}$) tidal streams observed in our M31 spectroscopic data are 0.08~dex and 0.32~dex, respectively (although the observed spread is likely underestimated due to systematic errors, \S\,\ref{sec:char_obs_sub}).  To compare the spread in [Fe/H] seen in the observations with that seen in the simulations, we impose surface brightness limits on the simulations based on the surface brightness of tidal streams observed in our survey (Fig.~\ref{fig:mufeh_data1}), and measure the rms devation of the mean [Fe/H] values of lines-of-sight through the BJ05 simulated stellar halo which have a stellar population dominated by one or two accretion events (open circles in Fig.~\ref{fig:sim_feh_alpha}).  The rms deviation of [Fe/H] values for high surface brightness lines-of-sight ($27<\mu_{\rm V}<29$~mag arcsec$^{-2}$) is 0.07~dex, while the rms deviation of [Fe/H] values for low surface brightness lines-of-sight ($29<\mu_{\rm V}<31.7$~mag arcsec$^{-2}$) is 0.13~dex.  The spread in [Fe/H] values in our observations and the simulations is similar for high surface brightness lines-of-sight, while the spread in [Fe/H] values in the observations is greater than in the simulations for lower surface brightness lines-of-sight.  The rms deviation of [$\alpha$/Fe] values in the simulations for the magnitude limits used above are 0.02~dex and 0.05 dex, respectively.

We note that the absolute [Fe/H] scale is uncertain for both the observations (due to the age-metallicity degeneracy of the RGB as well as systematic errors in the generation of theoretical isochrones, \S\,\ref{sec:data_met}) and the simulations (where the output [Fe/H] is dependent on the input chemical enrichment prescriptions, \S\,\ref{sec:sims_method}).  Therefore, we do not attempt to directly compare the metallicities of the data and simulations.  

\subsection{Discussion}\label{sec:comp_disc}


The common trend of increasing metallicity with increasing surface brightness in lines-of-sight through M31's stellar halo and simulated stellar halos is encouraging.  It gives us confidence that such simulations can lead to meaningful interpretation of stellar halo observations. 

The surface brightness of M31's stellar halo is observed to decrease with increasing distance from the center of the galaxy \citep{guhathakurta2005,irwin2005,ibata2007}, and the inner region of M31's stellar halo is observed to be more metal-rich than the outer region \citep{kalirai2006halo,chapman2006,koch2008}.  Thus, the general increase in mean [Fe/H] with increasing surface brightness of the kinematically broad components in our M31 data (Fig.~\ref{fig:mufeh_data1}, open squares) is expected.  In the hiearchical framework, this global trend in stellar halo metallicity could be explained with the same physical mechanisms invoked above (\S\,\ref{sec:comp_trends}).
Although the radial extent over which tidal debris from an individual accretion event is spread depends on the orbit of the progenitor, the effect of dynamical friction on the progenitor's orbit is more important for more massive progenitors, causing them to sink to the center of the gravitational potential of the host galaxy more quickly than less massive progenitors.  Therefore, more massive progenitors will tend to deposit most of their relatively metal-rich debris in the interior regions of the host halo.

If the number of M31 stars detected in an observational field is small ($\lesssim 20$), it can be difficult to 
determine if the field is dominated by contributions from one or two 
kinematical components or if the field 
contains contributions from multiple kinematical components (i.e., contains a relatively hot population).  
The results of the simulations can be used to examine the probable number of stellar streams
in a given field. If the mean \feh\ of a field is
relatively high compared to other fields of comparable surface brightness, the field may be dominated 
by one or two stellar streams, while if the mean \feh\ is relatively low, the field is more likely to be dominated by
multiple superposed stellar streams along the line of sight (i.e., the
stellar population along the line of sight is approaching a dynamically hot
population).  

There are several fields in our M31 spectroscopic survey in which we have
spectra of $\lesssim 10$ M31 RGB stars (open squares, Fig.~\ref{fig:mufeh_data2}); 
all are in M31's sparse outer halo ($R_{\rm proj}\gtrsim 80$~kpc).   
Based on the above discussion (\S\,\ref{sec:comp_trends}), the outer halo fields with the 
highest mean \feh\ values are the most likely to be dominated by debris from 
one or two high mass progenitors, while the fields with lower mean \feh\ are more likely to be 
dominated by debris from many lower mass progenitors.  Future observations will be able to 
test this conjecture.  If fields with lower mean \feh\ are found to have stars with a wide range of
velocities, it will support the hypothesis that the field is dominated by debris from many progenitors. 
Although the presence of a significant old, 
metal-poor stellar population in a higher mass progenitor may somewhat obscure this effect, 
the mean \feh\ of a field dominated by
such a progenitor would still be expected to be higher on average than the mean \feh\ of a field
dominated by low mass progenitors which did not have an appreciable number of metal-rich stars.
 
There are currently no observational constraints on the alpha-enrichment of 
stars in M31's stellar halo.  However, if a mean [$\alpha$/Fe] can be 
measured for the M31 RGB stars in a given field (i.e., by measuring the 
relative abundances of elements in a composite spectrum formed by coaddition 
of spectra from multiple stars), it could be used to constrain the average time 
of accretion of the stellar population.  Measurement of the mean [$\alpha$/Fe] would greatly facilitate 
the interpretation of the surface brightness of specific fields by lessening the degeneracy 
between the effect of the luminosity of accreted satellites versus the time of accretion onto the host halo.   

\section{Summary}\label{sec:sum}

We have presented the general trends of surface brightness with chemical enrichment of RGB stars observed in the course of our M31 Keck/DEIMOS spectroscopic survey and of stellar streams in simulations of stellar halos formed in a hierarchical framework.  Kinematically identified substructure in M31's stellar halo is metal-rich relative to the kinematically hot population, and higher surface brightness features are more metal-rich than lower surface brightness features.  

Similar trends are found in the simulated BJ05 stellar halos, which are formed entirely via accretion.  Higher surface brightness lines-of-sight through the simulated stellar halos are in general more metal-rich than lower surface brightness lines-of-sight, and lines-of-sight dominated by stars from one or two 
accretion events are in general more metal-rich than lines-of-sight whose stars come from many events.  The comparison of the observations and simulations provides a physical explanation for the observed trends in surface brightness and metallicity of stellar halo structure: more massive progenitors, which are also the most metal-rich, produce the brightest stellar halo substructures.  This results in the dominance of metal-rich streams in galactic halos.     

Both the data and the simulation results show a significant spread in metallicity at a given surface brightness for lines-of-sight through stellar halos.  This is a result of the varying time since accretion of the tidal streams' progenitors onto the host halo.  Tidal streams produced by more recent accretion events are, on average, brighter than tidal streams produced by less recent accretion events.  

The simulations show a relationship between the alpha-enhancement and surface brightness of tidal streams is expected: fainter tidal debris is on average more alpha-enhanced than brighter tidal debris.  This arises from the different times of accretion of progenitor satellites onto the host halo, as well as the assumption of the truncation of star formation upon accretion onto the host halo.  Satellites accreted early had less time for chemical enrichment via Type Ia SNe to reduce their alpha enrichment, while their tidal debris has had more time to spread along the progenitor's orbit within the host halo.  The ability to measure the alpha-enrichment of M31 RGB stars would thus lead to constraints on both the luminosity and the time of accretion onto the host halo of the progenitors of M31's tidal streams.


\acknowledgments 
The authors would like to thank James Bullock, Brant Robertson, Steve Majewski, Marla Geha, and Sanjib Sharma 
for their contributions to the research programs whose products were used 
in this work.  
The authors of the paper benefited from the hospitality of the Aspen Center for 
Physics in June 2006.  This work was supported by 
NSF grants AST-0307966, AST-0507483, and
AST-0607852 (K.M.G., P.G.) and a UCSC Chancellor's Division Dissertation-Year Fellowship (K.M.G.), NSF CAREER award AST-0733966 and NSF grant AST-0734864 (K.V.J.), and an STFC  Fellowship at the Institute for Computational Cosmology in Durham (A.S.F.).


\bibliography{m31}




\end{document}